# Generalized Finsler geometry and its Cartan connection in modification of special relativity

Jian-Miin Liu*
Department of Physics, Nanjing University
Nanjing, The People's Republic of China
* on leave, e-mail: liu@phys.uri.edu

ABSTRACT
   The generalized Finsler geometry, as well as Finsler geometry, is a generalization of Riemann geometry. The generalized Finsler geometry can be endowed with the Cartan connection. The generalized Finsler geometry and its Cartan connection are, as the necessary mathematical tools, involved in our modification of special relativity, which is made by assuming the generalized Finslerian structures of gravity-free space and time in the usual inertial coordinate system and combining this assumption with two fundamental postulates, (i) the principle of relativity and (ii) the constancy of the speed of light in all inertial frames of reference.

## 1. INTRODUCTION

Many relevant books and articles begin with a "taken" inertial coordinate system that we call the usual inertial coordinate system and denote with $\{x^r, t\}$, $r=1,2,3$, or $\{x^i\}$, $i=1,2,3,4$, where $x^4=ict$ and $c$ is the speed of light. In it the authors of these books and articles assume the Euclidean structures of gravity-free space and time,

$$dX^2 = \delta_{rs} dx^r dx^s, \quad r,s=1,2,3, \tag{1a}$$
$$dT^2 = dt^2, \tag{1b}$$

or, equivalently in the four-dimensional form,

$$d\Sigma^2 = \delta_{ij} dx^i dx^j, \quad i,j=1,2,3,4, \tag{2}$$

and write down various equations based on several basic principles on symmetries. They then do calculations, obtain results and compare the obtained results to experimental measurements. By "taken", we mean only calling and using without discussing this coordinate system. There are at least two important questions about it. The first question is whether the measurement method associated with the usual inertial coordinate system is that we use in our experiments? In other words, what is the definition of the usual inertial coordinate system from the measurement point of view? The second one is whether or not the local structures of gravity-free space and time are Euclidean in the usual inertial coordinate system?

Carrying out researches in the field of theoretical physics, we seem to have to answer the first question before we compare those obtained results to experiments and be sure with the local structures of gravity-free space and time before we write down equations and do calculations.

Einstein exquisitely answered the first question. Conceptually, there is a class of equivalent inertial frames of reference, any one of which moves with a non-zero constant velocity relative to any other and is supplied with motionless, rigid unit rods of equal length and motionless, synchronized clocks of equal running rate. For each inertial frame of reference, Einstein defined the usual inertial coordinate system: "in a given inertial frame of reference the coordinates mean the results of certain measurements with rigid (motionless) rods, a clock at rest relative to the inertial frame of reference defines a local time, and the local time at all points of space, indicated by synchronized clocks and taken together, give the time of this inertial frame of reference."[1] As defined, in each inertial frame of reference, an observer who is at rest with respect to that inertial frame of reference can employ his own motionless-rigid rods and motionless-synchronized clocks in the so-called "motionless-rigid rod and motionless-synchronized clock" measurement method to measure space and time intervals. By using this measurement method, he sets up his usual inertial coordinate system. Since the "motionless-rigid rod and motionless-synchronized clock" measurement method is that we use in our experiments, it is in reason to compare the calculation results obtained in the usual inertial coordinate system to experimental measurements.

To the second question we recently gave a negative answer [2]. The generalized Finslerian structures of gravity-free space and time in the usual inertial coordinate system were assumed [3,4].

Combining this assumption with two fundamental postulates on the principle of relativity and the constancy of the light speed, we modified special relativity. We re-formulated mechanics, field theory and quantum field theory. The generalized Finsler geometry and its Cartan connection, as the necessary mathematical tools, were involved in.

2. FINSLER GEOMETRY

Finsler geometry is a generalization of Riemann geometry. It was first suggested by Riemann as early as 1854 [5], and studied systematically by Finsler in 1918 [6]. In Finsler geometry [5-10], distance ds between two neighboring points $x^k$ and $x^k+dx^k$, k=1,2,---,n is defined by a scale function

$$ds = F(x^1, x^2, ---, x^n; dx^1, dx^2, ---dx^n)$$

or simply

$$ds = F(x^k, dx^k), \quad k=1,2,---,n, \tag{3}$$

which depends on directional variables $dx^k$ as well as coordinate variables $x^k$. Apart from several routine conditions like smoothness, the main constrains imposed on $F(x^k, dx^k)$ are:

(A) It is positively homogeneous of degree one in $dx^k$,

$$F(x^k, \lambda dx^k) = \lambda F(x^k, dx^k) \quad \text{for } \lambda > 0; \tag{4}$$

(B) It is positive if not all $dx^k$ are simultaneously zero,

$$F(x^k, dx^k) > 0 \quad \text{for } dx^k dx^k \neq 0; \tag{5}$$

(C) The quadratic form

$$\frac{\partial^2 F^2(x^k, dx^k)}{\partial dx^i \partial dx^j} P^i P^j, \quad i,j=1,2,---,n, \tag{6}$$

is positive definite for all vector $P^i$ and any $(x^k, dx^k)$.

Introducing a set of equations

$$g_{ij}(x^k, dx^k) = \frac{1}{2} \frac{\partial^2 F^2(x^k, dx^k)}{\partial dx^i \partial dx^j}, \quad i,j=1,2,---,n, \tag{7}$$

we can represent Finsler geometry in terms of

$$ds^2 = g_{ij}(x^k, dx^k) dx^i dx^j, \tag{8}$$

where $g_{ij}(x^k, dx^k)$ is called the Finslerian metric tensor induced from scale function $F(x^k, dx^k)$. The Finslerian metric tensor is symmetric in its subscripts and all its components are positively homogeneous of degree zero in $dx^k$,

$$g_{ij}(x^k, dx^k) = g_{ji}(x^k, dx^k), \tag{9}$$

$$g_{ij}(x^k, \lambda dx^k) = g_{ij}(x^k, dx^k) \quad \text{for } \lambda > 0. \tag{10}$$

The Riemann geometry with metric tensor $g_{ij}(x^k)$ is nothing but the Finsler geometry with scale function $F(x^k, dx^k) = [g_{ij}(x^k) dx^i dx^j]^{1/2}$.

3. GENERALIZED FINSLER GEOMETRY

We can define the so-called generalized Finsler geometry by omitting Eq.(3) as a definition of ds and instead taking Eq.(8) as a definition of $ds^2$, where metric tensor $g_{ij}(x^k, dx^k)$ is given and satisfies, besides Eqs.(9) and (10) and some routine conditions, that quadratic form

$$g_{ij}(x^k, dx^k) P^i P^j, \quad i,j=1,2,---,n, \tag{11}$$

is positive definite for all $P^i$ and any $(x^k, dx^k)$.

The generalized Finsler geometry is so-named because a Finsler geometry must be a generalized Finsler geometry while a generalized Finsler geometry is not necessarily a Finsler geometry [10].

To see this, we prove a statement: The generalized Finsler geometry with positive definite metric $g_{ij}(x^k, dx^k)$ conditioned by Eqs.(9) and (10) is a Finsler geometry when and only when we can find a scale function $F(x^k, dx^k)$ by solving the set of Eqs.(7) such that this function satisfies conditions (A)-(C).

Actually, if this generalized Finsler geometry is a Finsler geometry, there must be a scale function $F'(x^k, dx^k)$ satisfying conditions (A)-(C) and $ds = F'(x^k, dx^k)$. From $F'(x^k, dx^k)$, we can find

$$g'_{ij}(x^k, dx^k) = \partial^2 F'^2(x^k, dx^k)/2 \partial dx^i \partial dx^j, \tag{12}$$

and represent the Finsler geometry as

$$ds^2 = g'_{ij}(x^k, dx^k) dx^i dx^j.$$

This equation coincides with Eq.(8) because the Finsler geometry is originally the generalized Finsler geometry with metric $g_{ij}(x^k,dx^k)$, in other words, $g'_{ij}(x^k,dx^k)=g_{ij}(x^k,dx^k)$. Substituting $g_{ij}(x^k,dx^k)$ for $g'_{ij}(x^k,dx^k)$ in Eq.(12), we find scale function $F'(x^k,dx^k)$ being such a solution to the set of Eqs.(7). Reversely, if for this generalized Finsler geometry we can solve the set of Eqs.(7) and find a scale function $F(x^k,dx^k)$ satisfying conditions (A)-(C), $F^2(x^k,dx^k)$ must be positively homogeneous of degree two in $dx^k$. The following equation holds due to Euler's theorem on homogeneous functions,

$$2F^2(x^k,dx^k)= dx^i [\partial F^2(x^k,dx^k)/\partial dx^i]. \quad (13)$$

Iterating Eq.(13), we have

$$4F^2(x^k,dx^k)= dx^j \delta_{ij}[\partial F^2(x^k,dx^k)/\partial dx^i]$$
$$+dx^j dx^i[\partial^2 F^2(x^k,dx^k)/\partial dx^j \partial dx^i]. \quad (14)$$

Using Eq.(13) in Eq.(14), we have further

$$F^2(x^k,dx^k)= dx^j dx^i[\partial^2 F^2(x^k,dx^k)/2\partial dx^j \partial dx^i]. \quad (15)$$

Eq.(15) combines with Eqs.(7) and (8) to yield Eq.(3). This generalized Finsler geometry is a Finsler geometry.

As the set of partial differential equations, Eqs.(7), does not always have such solution of scale function $F(x^k,dx^k)$ for an arbitrarily given positive definite metric $g_{ij}(x^k,dx^k)$ conditioned by Eqs.(9-10), a generalized Finsler geometry is not necessarily a Finsler geometry.

4. THE CARTAN CONNECTION

The generalized Finsler geometry can be endowed with the Cartan connection. There are two kinds of covariant partial derivative. There exist three distinct curvature tensors that we call the Cartan curvature tensors.

4.1. The Cartan connection

Let us consider the generalized Finsler geometry with metric tensor $g_{ij}(x^k,dx^k)$ or $g_{ij}(x^k,z^k)$, where $z^k=dx^k/db$, $k=1,2,\cdots,n$, and $db$ is a positive invariant. The pair $(x^k,z^k)$ is called the element of support.

According to Cartan, the absolute differentials of vectors $X^i$ and $X_i$ can be written as

$$DX^i=dX^i+ C^i_{kh}(x,z)X^k dz^h + \Gamma^i_{kh}(x,z)X^k dx^h, \quad (16a)$$
$$DX_i=dX_i - C^k_{ih}(x,z)X_k dz^h - \Gamma^k_{ih}(x,z)X_k dx^h, \quad (16b)$$

where $C^i_{kh}$ and $\Gamma^i_{kh}$ are the affine coefficients. In Eqs.(16a-16b) abbreviations x and z are respectively taken for $x^k$ and $z^k$ in parentheses. Two kinds of covariant partial derivative, $X^i_{|h}$, $X_{i|h}$ and $X^i_{\|h}$, $X_{i\|h}$, can be defined by

$$DX^i= X^i_{|h}dx^h + X^i_{\|h}Dz^h, \quad (17a)$$
$$DX_i= X_{i|h}dx^h + X_{i\|h}Dz^h, \quad (17b)$$

where $Dz^k$ denotes the absolute differential of $z^k$.

Again according to Cartan, the following conditions determine the Cartan connection.

(a) If the direction of vector $P^i$ coincides with that of its element of support (x,z), its square of length is to be equal to

$$g_{ij}(x,P)P^i P^j. \quad (18a)$$

(b) If $P^i$ and $Q^i$ represent two vectors with a common element of support (x,z), a symmetry stands between them,

$$g_{ij}(x,z)Q^i(DP^j)=g_{ij}(x,z)P^i(DQ^j). \quad (18b)$$

(c) for vector $z^k$, it holds that

$$C^k_{ij}(x,z)z^i dz^j=0, \quad k,i,j=1,2,\cdots,n. \quad (18c)$$

(d) The coefficients $\Gamma^{*i}_{hk}(x,z)$, whose definitions are given below, are symmetric in their lower indices,

$$\Gamma^{*i}_{hk}(x,z)= \Gamma^{*i}_{kh}(x,z). \quad (18d)$$

In the Cartan connection, we have

$$\nabla_h X^i \equiv X^i_{|h}= \partial_h X^i - (\partial G^k/\partial z^h)(\partial X^i/\partial z^k) + \Gamma^{*i}_{kh}X^k, \quad (19a)$$
$$\nabla_h X_i \equiv X_{i|h}= \partial_h X_i - (\partial G^k/\partial z^h)(\partial X_i/\partial z^k) - \Gamma^{*k}_{ih}X_k, \quad (19b)$$
$$X^i_{\|h}= \partial X^i/\partial z^h + C^i_{kh}X^k, \quad (20a)$$
$$X_{i\|h}= \partial X_i/\partial z^h - C^k_{ih}X_k, \quad (20b)$$

where

$$\Gamma^{*k}_{ih} = \Gamma^k_{ih} - C^k_{im}(\partial G^m/\partial z^h), \tag{21a}$$

$$C^k_{jh} = g^{km} C_{jmh}, \tag{21b}$$

$$\Gamma^k_{jh} = g^{km} \Gamma_{jmh} = g^{km}\{\gamma_{jmh} - C_{hmp}(\partial G^p/\partial z^j) + C_{jhp}(\partial G^p/\partial z^m)\}, \tag{21c}$$

$$\gamma_{jkh} = \frac{1}{2}\{\partial g_{jk}/\partial x^h + \partial g_{kh}/\partial x^j - \partial g_{hj}/\partial x^k\}, \tag{21d}$$

$$G^p = \frac{1}{2}\gamma^p_{hj} z^h z^j, \tag{21e}$$

$$\gamma^p_{hj} = g^{pk}\gamma_{hkj}, \tag{21f}$$

$$C_{jmh} = \frac{1}{2}(\partial g_{jm}/\partial z^h). \tag{21g}$$

As for tensor $T^i_j$ and scale function $Q(x,z)$, two kinds of covariant partial derivative are respectively,

$$\nabla_h T^i_j \equiv T^i_{j|h} = \partial_h T^i_j - (\partial G^k/\partial z^h)(\partial T^i_j/\partial z^k) + \Gamma^{*i}_{kh}T^k_j - \Gamma^{*k}_{jh}T^i_k, \tag{22}$$

$$T^i_{j\|h} = \partial T^i_j/\partial z^h + C^i_{kh}T^k_j - C^k_{jh}T^i_k, \tag{23}$$

and

$$\nabla_h Q \equiv Q_{|h} = \partial_h Q - (\partial G^k/\partial z^h)(\partial Q/\partial z^k), \tag{24}$$

$$Q_{\|h} = \partial Q/\partial z^h. \tag{25}$$

Especially, for metric tensor $g_{ij}(x,z)$, we have

$$Dg_{ij} = 0, \tag{26a}$$

$$g_{ij|h} = 0, \tag{26b}$$

$$g_{ij\|h} = 0. \tag{26c}$$

4.2. The Cartan curvature tensors

In the generalized Finsler geometry, there are two distinct kinds of covariant partial derivative for vector $X^i$, $X^i_{|h}$ (the first kind) and $X^i_{\|h}$ (the second kind), and hence, three distinct kinds of second-order covariant partial derivatives, $X^i_{|h|k}$, $X^i_{|h\|k}$ and $X^i_{\|h\|k}$. Consequently, there are three distinct curvature tensors which are respectively in connection with commutation relations, $X^i_{|h|k} - X^i_{|k|h}$, $X^i_{|h\|k} - X^i_{\|k|h}$ and $X^i_{\|h\|k} - X^i_{\|k\|h}$,

$$R^i_{jkh} = \{\partial \Gamma^{*i}_{jh}/\partial x^k - (\partial \Gamma^{*i}_{jh}/\partial z^m)(\partial G^m/\partial z^k)\} - \{\partial \Gamma^{*i}_{jk}/\partial x^h - (\partial \Gamma^{*i}_{jk}/\partial z^m)(\partial G^m/\partial z^h)\}$$
$$+ C^i_{jm}\{\partial^2 G^m/\partial x^k \partial z^h - \partial^2 G^m/\partial z^k \partial x^h - G^m_{hn}(\partial G^n/\partial z^k) + G^m_{kn}(\partial G^n/\partial z^h)\}$$
$$+ \Gamma^{*i}_{mk}\Gamma^{*m}_{jh} - \Gamma^{*i}_{mh}\Gamma^{*m}_{jk}, \tag{27}$$

$$P^i_{jkh} = (\partial \Gamma^{*i}_{jk}/\partial z^h) + C^i_{jm}C^m_{hk|r}z^r - C^i_{jh|k} \tag{28}$$

and

$$S^i_{jkh} = C^i_{kr}C^r_{jh} - C^i_{rh}C^r_{jk}, \tag{29}$$

where

$$G^i_{hk} = \partial^2 G^i/\partial z^h \partial z^k. \tag{30}$$

These curvature tensors satisfy a large number of identities.

5. SPECIAL CASES OF GENERALIZED FINSLER GEOMETRY

If metric tensor $g_{ij}(x^k,z^k)$ is $z^k$-independent, Eq.(8) becomes

$$ds^2 = g_{ij}(x^k)dx^i dx^j. \tag{31}$$

In the case, the generalized Finsler geometry reduces to a Riemannian geometry and the Cartan connection reduces to the Levi-Civita connection. Moreover, tensor $R^i_{jkh}$ reduces to the Riemann curvature tensor, and

$$S^i_{jkh} = P^i_{jkh} = 0. \tag{32}$$

If metric tensor $g_{ij}(x^k,z^k)$ is $x^k$-independent, the covariant partial derivative of the first kind reduces to the usual partial derivative, namely

$$\nabla_h = \partial_h, \tag{33}$$

and

$$R^i_{jkh} = 0. \tag{34}$$

# 6. LOCAL STRUCTURES OF GRAVITY-FREE SPACE AND TIME

Einstein defined the usual inertial coordinate system $\{x^r,t\}$, $r=1,2,3$, for each inertial frame of reference, through associating it with the "motionless-rigid rod and motionless-synchronized clock" measurement method. The "motionless-rigid rod and motionless-synchronized clock" measurement method is not the only one that the motionless observer in each inertial coordinate system has. We imagine other measurement methods for him. By taking these other measurement methods he can set up other inertial coordinate systems, just as well as he can set up the usual inertial coordinate system by taking the "motionless-rigid rod and motionless-synchronized clock" measurement method. One of these other inertial coordinate systems is called the primed inertial coordinate system and denoted with $\{x'^r,t'\}$, $r=1,2,3$.

In Ref.[3,4], we assumed that gravity-free space and time possess the Euclidean structures in the primed inertial coordinate system and the following non-Euclidean structures in the usual inertial coordinate system,

$$dX^2 = \delta_{rs} dx'^r dx'^s = g_{rs}(dx^1,dx^2,dx^3,dt) dx^r dx^s, \quad r,s=1,2,3, \tag{35a}$$
$$dT^2 = dt'^2 = g(dx^1,dx^2,dx^3,dt) dt^2, \tag{35b}$$
$$g_{rs}(dx^1,dx^2,dx^3,dt) = K^2(y)\delta_{rs}, \tag{35c}$$
$$g(dx^1,dx^2,dx^3,dt) = (1-y^2/c^2), \tag{35d}$$
$$K(y) = \frac{c}{2y}(1-y^2/c^2)^{1/2} \ln\frac{c+y}{c-y}, \tag{35e}$$
$$y = (y^s y^s)^{1/2}, \tag{35f}$$
$$y^s = dx^s/dt, \tag{35g}$$

where $dX$ and $dT$ are respectively the real distance and time differentials between two neighboring points $(x'^1,x'^2,x'^3,t')$ and $(x'^1+dx'^1,x'^2+dx'^2,x'^3+dx'^3,t'+dt')$ in the primed inertial coordinate system or $(x^1,x^2,x^3,t)$ and $(x^1+dx^1,x^2+dx^2,x^3+dx^3,t+dt)$ in the usual inertial coordinate system. The generalized Finslerian structures of gravity-free space and time in the usual inertial coordinate system are specified by metric tensors $g_{rs}(dx^1,dx^2,dx^3,dt)$ and $g(dx^1,dx^2,dx^3,dt)$, which depend only on directional variables and become Euclidean when and only when y approaches zero.

We also combined the assumption Eqs.(35) with two fundamental postulates, (i) the principle of relativity and (ii) the constancy of the speed of light in all inertial frames of reference, to modify special relativity. The modified special relativity theory involves two versions of the light speed, infinite

$$c' = \lim_{y \to c} \frac{c}{2} \ln\frac{c+y}{c-y} \tag{36}$$

in the primed inertial coordinate system and finite c in the usual inertial coordinate system. It involves the c'-type Galilean transformation between any two primed inertial coordinate systems and the localized Lorentz transformation between any two usual inertial coordinate systems. Since all our experimental data are collected and expressed in the usual inertial coordinate system, the physical principle in the modified special relativity theory is: the c'-type Galilean invariance in the primed inertial coordinate system plus the transformation from the primed inertial coordinate system to the usual inertial coordinate system. In the primed inertial coordinate system, we write all physical laws in the c'-type Galilean-invariant form and do all calculations in the c'-type Galilean-invariant manner; we finally transform these calculation results from the primed inertial coordinate system to the usual inertial coordinate system and compare them to experimental facts. The assumption Eqs.(35) implies

$$Y^2 = \delta_{rs} y'^r y'^s = [\frac{c}{2y} \ln\frac{c+y}{c-y}]^2 \delta_{rs} y^r y^s, \quad r,s=1,2,3, \tag{37}$$

which embodies the velocity space defined by

$$dY^2 = \delta_{rs} dy'^r dy'^s, \quad r,s=1,2,3, \tag{38a}$$

in the primed velocity-coordinates $\{y'^r\}$, $r=1,2,3$, or

$$dY^2 = H_{rs}(y) dy^r dy^s, \quad r,s=1,2,3, \tag{38b}$$
$$H_{rs}(y) = c^2 \delta^{rs}/(c^2-y^2) + c^2 y^r y^s/(c^2-y^2)^2, \quad \text{real } y^r \text{ and } y<c, \tag{38c}$$

in the usual (Newtonian) velocity-coordinates {$y^r$}, r=1,2,3, where Y=dX/dT is the real speed or velocity-length, $y'^r=dx'^r/dt'$ is called the primed velocity, $y^r=dx^r/dt$ is the well-defined Newtonian velocity. The velocity space has the Riemannian structure in the usual velocity-coordinates. The different representations of the velocity space in the primed and the usual velocity-coordinates led to the relativistic corrections to the Maxwellian velocity distribution [11].

It is very important to define the primed inertial coordinate system from the measurement point of view. We do this somewhere else [12].

ACKNOWLEDGMENT

The author greatly appreciates the teachings of Prof. Wo-Te Shen. The author thanks Prof. John S. Desjardins for his suggestions and supports of this work.